# Similarity for ultra-relativistic laser plasmas and the optimal acceleration regime

S. Gordienko[1,2], A. Pukhov[1]

[1]*Institut für Theoretische Physik I, Heinrich-Heine-Universität Düsseldorf, D-40225, Germany*
[2]*L. D. Landau Institute for Theoretical Physics, Moscow, Russia*
(Dated: November 9, 2004)

A similarity theory is developed for ultra-relativistic laser-plasmas. It is shown that the most fundamental $S-$similarity is valid for both under- and overdense plasmas. Optimal scalings for laser wake field electron acceleration are obtained *heuristically*. The strong message of the present work is that the bubble acceleration regime [see Pukhov, Meyer-ter-Vehn, Appl. Phys. B, **74**, 355 (2002)] satisfies these optimal scalings.

The concept of laser-plasma electron acceleration has the decisive advantage over conventional accelerators: plasma supports electric fields orders of magnitude higher than the breakdown-limited field in radio-frequency cavities of conventional linacs. It is expected that the relativistic laser-plasma will finally lead to a compact high energy accelerator [1]. The very first experiments already have delivered high quality electron beams in the energy range 70...170 MeV [2–4]. Yet the way to a real laser-plasma accelerator that generates a high-energy electron beam with parameters required by practical applications is long and full of problems which have to be solved. The main obstacle is that the experiments depend on too many parameters. Often, this makes the interpretation of experimental results ambiguous. At the same time, theoretical models suffer from the similar drawback. The system of kinetic equations describing the problem is both strongly non-linear and contains many parameters. As a result, the quest of searching for new perspective acceleration regimes is challenging and the physics of electron acceleration in plasma is often rather obscure.

The scientific difficulties just listed are neither new nor unique. Quite analogous problems encounters the classical (magneto-)hydrodynamics. One of the most powerful theoretical tools in such situation is the similarity theory [5, 6]. The similarity allows engineers to scale the behavior of a physical system from a laboratory acceptable size to the size of practical use.

To the best of our knowledge, no similarity theory has been applied to relativistic laser plasma interactions. This situation is surprising and unnatural, because the power of similarity theory for the magnetic confinement was recognized in the late 70s and the similarity theory [7] has been in use for design of large devices (tokamaks, stellarators) ever thereafter [8].

For the first time, we develop a similarity theory for laser-plasma interactions in the ultra-relativistic limit. Using a fully kinetic approach we show that the similarity parameter $S = n_e/a_0 n_c$ exists, where $a_0 = eA_0/m_e c^2$ is the relativistically normalized laser amplitude, $n_e$ is the plasma electron density and $n_c = m_e \omega_0^2/4\pi e^2$ is the critical density for a laser with the frequency $\omega_0$. The basic ultra-relativistic similarity states that laser-plasma interactions with different $a_0$ and $n_e/n_c$ are similar as soon as the parameter $S = n_e/a_0 n_c = const$.

The basic $S-$similarity is valid for both over- and underdense plasmas. In the present work, we are interested in the special limit $S \ll 1$ of relativistically underdense plasmas as it is important for the high energy electron acceleration. In this case, $S$ can be considered as a small parameter and quite general scalings for laser-plasma interactions can be found. It follows from the theory that in the optimal configuration the laser pulse has the focal spot radius $k_p R \approx \sqrt{a_0}$ and the duration $\tau \leq R/c$. Here, $k_p = \omega_p/c$ is the plasma wavenumber and $\omega_p^2 = 4\pi n_e e^2/m_e$ is the plasma frequency. This corresponds to the "Bubble" acceleration regime [9].

The central result of our work is that the bubble regime satisfies the optimal wake field acceleration scalings. The scaling for the maximum energy $E_{mono}$ of the monoenergetic peak in the electron spectrum is

$$E_{mono} \approx 0.65 m_e c^2 \sqrt{\frac{\mathcal{P}}{\mathcal{P}_{rel}}} \frac{c\tau}{\lambda}. \qquad (1)$$

Here, $\mathcal{P}$ is the laser pulse power, $\mathcal{P}_{rel} = m_e^2 c^5/e^2 \approx 8.5$ GW is the natural relativistic power unit, and $\lambda = 2\pi c/\omega_0$ is the laser wavelength. The scaling (1) assumes that the laser pulse duration satisfies the condition $c\tau < R$. The scaling for the number of accelerated electrons $N_{mono}$ in the monoenergetic peak is

$$N_{mono} \approx \frac{1.8}{k_0 r_e} \sqrt{\frac{\mathcal{P}}{\mathcal{P}_{rel}}}, \qquad (2)$$

where $r_e = e^2/m_e c^2$ is the classical electron radius, and $k_0 = 2\pi/\lambda$. The acceleration length $L_{acc}$ scales as

$$L_{acc} \approx 0.7 \frac{c\tau}{\lambda} Z_R, \qquad (3)$$

where $Z_R = \pi R^2/\lambda \approx a_0 \lambda_p^2/4\pi\lambda$ is the Rayleigh length.

The parametric dependencies in the scalings (1)-(3) follow from the analytical theory. The numerical prefactors are taken from 3D PIC simulations.

We consider collisionless laser-plasma dynamics and neglect the ion motion. The electron distribution function $f(t, \mathbf{r}, \mathbf{p})$ is described by the Vlasov equation

$$(\partial_t + \mathbf{v}\partial_\mathbf{r} - e(\mathbf{E} + \mathbf{v}\times\mathbf{B}/c)\partial_\mathbf{p}) f(t,\mathbf{p},\mathbf{r}) = 0, \quad (4)$$

where $\mathbf{p} = m_e\gamma\mathbf{v}$ and self-consistent fields $\mathbf{E}$ and $\mathbf{B}$ satisfy the Maxwell equations [11].

We suppose that the laser pulse vector potential at the time $t=0$ short before entering the plasma is $\mathbf{A}(t=0) = \mathbf{a}\left((y^2+z^2)/R^2, x/c\tau\right)\cos(k_0 x)$, where $k_0 = \omega_0/c$ is the wavenumber, $R$ is the focal spot radius and $\tau$ is the pulse duration. If one fixes the laser envelope $\mathbf{a}(\mathbf{r}_\perp, x)$, then the laser-plasma dynamics depends on four dimensionless parameters: the laser amplitude $a_0 = \max|e\mathbf{a}/mc^2|$, the focal spot radius $k_0 R$, the pulse duration $\omega_0\tau$, and the plasma density ratio $n_e/n_c$, where $n_c = m_e\omega_0^2/4\pi e^2$ is the critical density.

Now we are going to show that in the ultra-relativistic limit when $a_0 \gg 1$, the number of independent dimensionless parameters reduces to three: $k_0 R$, $\omega_0\tau$ and $S$, where the similarity parameter $S$ is

$$S = \frac{n_e}{a_0 n_c}. \quad (5)$$

Let us introduce the new dimensionless variables

$$\hat{t} = S^{1/2}\omega_0 t, \quad \hat{\mathbf{r}} = S^{1/2}k_0\mathbf{r}, \quad \hat{\mathbf{p}} = \mathbf{p}/m_e c a_0, \quad (6)$$
$$\hat{\mathbf{A}} = \frac{e\mathbf{A}}{mc^2 a_0}, \quad \hat{\mathbf{E}} = \frac{S^{-1/2}e\mathbf{E}}{mc\omega_0 a_0}, \quad \hat{\mathbf{B}} = \frac{S^{-1/2}e\mathbf{B}}{mc\omega_0 a_0},$$

and the new distribution function $\hat{f}$ defined as

$$f = \frac{n_e}{(m_e c a_0)^3}\hat{f}\left(\hat{t}, \hat{\mathbf{p}}, \hat{\mathbf{r}}, a_0, S, \hat{R}, \hat{\tau}\right), \quad (7)$$

where $\hat{R} = S^{1/2}k_0 R$ and $\hat{\tau} = S^{1/2}\omega_0\tau$.

The normalized distribution function $\hat{f}$ is a universal one describing the interaction of the given laser pulse with a fixed plasma profile. It satisfies the equations

$$\left[\partial_{\hat{t}} + \hat{\mathbf{v}}\partial_{\hat{\mathbf{r}}} - \left(\hat{\mathbf{E}} + (\hat{\mathbf{v}}\times\hat{\mathbf{B}})\partial_{\hat{\mathbf{p}}}\right)\right]\hat{f} = 0, \quad (8)$$
$$\nabla_{\hat{\mathbf{r}}}\cdot\hat{\mathbf{E}} = 4\pi(1+\hat{\rho}), \quad \nabla_{\hat{\mathbf{r}}}\cdot\hat{\mathbf{B}} = 0, \quad (9)$$
$$\nabla_{\hat{\mathbf{r}}}\times\hat{\mathbf{B}} = 4\pi\hat{\mathbf{j}} + \partial_{\hat{t}}\hat{\mathbf{E}}, \quad \nabla_{\hat{\mathbf{r}}}\times\hat{\mathbf{E}} = -\partial_{\hat{t}}\hat{\mathbf{B}},$$

where $\hat{\mathbf{v}} = \hat{\mathbf{p}}/\sqrt{\hat{\mathbf{p}}^2 + a_0^{-2}}$, $\hat{\rho} = -\int\hat{f}\,d\hat{\mathbf{p}}$, $\hat{\mathbf{j}} = -\int\hat{\mathbf{v}}\hat{f}\,d\hat{\mathbf{p}}$ and the initial condition for the vector potential is

$$\hat{\mathbf{A}}(\hat{t}=0) = \hat{\mathbf{a}}\left((\hat{y}^2+\hat{z}^2)/\hat{R}, \hat{x}/\hat{\tau}\right)\cos\left(S^{-1/2}\hat{x}\right), \quad (10)$$

with the slow envelope $\hat{\mathbf{a}}$ such that $\max|\hat{\mathbf{a}}| = 1$.

Eqs. (8) together with the initial condition (10) still depend on the four dimensionless parameters $\hat{R}, \hat{\tau}, S$ and $a_0$. However, the parameter $a_0$ appears only in the expression for the electron velocity $\hat{\mathbf{v}} = \hat{\mathbf{p}}/\sqrt{\hat{\mathbf{p}}^2 + a_0^{-2}}$. In the limit $a_0 \gg 1$ one can write $\hat{\mathbf{v}} = \hat{\mathbf{p}}/|\hat{\mathbf{p}}|$. In this limit, the laser-plasma dynamics does not depend separately on $a_0$ and $n_e/n_c$. Rather, they converge into the single similarity parameter $S$.

The ultra-relativistic similarity means that for different interaction cases with $S = const$, plasma electrons move along similar trajectories. Number of these electron $N_e$, their momenta $\mathbf{p}$, and the plasma fields scale as

$$\mathbf{p} \propto a_0; \quad N_e \propto a_0; \quad (11)$$
$$\phi, \mathbf{A}, \mathbf{E}, \mathbf{B} \propto a_0 \quad (12)$$

for $\omega_0\tau = const$, $k_0 R = const$ and $S = const$.

The ultra-relativistic similarity is valid for arbitrary $S$-values. The $S$ parameter appears only in the initial condition (10) so that $S^{-1/2}$ plays the role of the laser frequency. It separates the relativistically overdense plasmas with $S \gg 1$ from the underdense ones with $S \ll 1$.

From now on, we concentrate on the special case of underdense plasma, $S \ll 1$. In this paper, we provide a *heuristic* derivation of scalings for the optimal regime of wake field acceleration. A mathematically accurate and detailed derivation will be published elsewhere [12].

If we fix the laser pulse envelope, then the laser-plasma dynamics depends on the three dimensionless parameters: the pulse radius $k_0 R$, its aspect ratio $\Pi = c\tau/R$ and the parameter $S$. In the case of tenious plasmas, $S \ll 1$, one can develop all the universal functions around $S \approx 0$ and obtain additional similarities. An additional similarity means that the number of truly independent dimensionless parameters decreases, i.e., one of the parameters $S, k_0 R$, or $\Pi$ can be expressed as a function of the remaining two. We choose $S$ and $\Pi$ as the independent parameters:

$$\gamma = S^{\alpha_\gamma}a_0\gamma_0(\Pi), \quad N = S^{\alpha_N}a_0 N_0(\Pi), \quad (13)$$
$$R = S^{\alpha_R}R_0(\Pi), \quad L = S^{\alpha_L}L_0(\Pi),$$

where $\alpha_i$ are unknown powers and $\gamma_0(\Pi), N_0(\Pi), R_0(\Pi), L_0(\Pi))$ depend on the only dimensionless parameter $\Pi$. In this notations, $N$ is the number of trapped electrons and $L$ is the acceleration length.

It follows from the Maxwell equations that the maximum possible accelerating wake field scales as $E_{wake} \propto n_e R$. It corresponds to complete electron expulsion by the laser pulse ponderomotive pressure. This field accelerates electrons to the energy

$$m_e c^2 \gamma = \varkappa e L n_e R, \quad (14)$$

where $\varkappa$ is the acceleration efficiency.

The laser energy is deposited in plasma in form of the field $E_{wake}$. We introduce the laser depletion factor $\theta$:

$$E_{wake}^2 R^2 L = \theta W_{laser}, \qquad (15)$$

where $W_{laser} m c^2 n_c a_0^2 R^3 \Pi$ is the laser pulse energy. Analogously, energy of the electron bunch is

$$m_e c^2 \gamma N_{tr} = \eta W_{laser}, \qquad (16)$$

where $\eta$ is the overall energy conversion efficiency.

The similarity demands that

$$\varkappa = S^{\alpha_\varkappa} \varkappa_0(\Pi); \quad \theta = S^{\alpha_\theta} \theta_0(\Pi); \quad \eta = S^{\alpha_\eta} \eta_0(\Pi). \quad (17)$$

Because of their physical meaning, the energy efficiencies satisfy the inequalities $0 \leq \eta \leq \theta \leq 1$. To ensure scalability towards $S \ll 1$, the conditions $\alpha_\eta \geq \alpha_\theta \geq 0$ must be satisfied. Also, the accelerating rate (14) cannot be parametrically larger than that defined by $E_{wake}$. Consequently, we have $\alpha_\varkappa \geq 0$.

From now on, we demand for the most efficient scalable regime and set $\alpha_\varkappa = \alpha_\eta = \alpha_\theta = 0$. Eqs. (13)-(17) relate the unknown powers $\alpha_i$ and the $\Pi-$dependent functions:

$$\alpha_\gamma = \alpha_R + \alpha_L + 1, \quad \alpha_\gamma = 3\alpha_R - \alpha_N, \quad (18)$$
$$-2 = \alpha_R + \alpha_L, \quad (19)$$
$$\gamma_0 = \varkappa_0 k_0^2 R_0 L_0, \quad \gamma_0 N_0 = \Pi \theta_0, \quad k_0^2 R_0 L_0 = \Pi \theta_0. \quad (20)$$

The three equations (18)-(19) contain four unknown variables and, generally speaking, are insufficient to define *all* the powers $\alpha_i$. This is not unexpected, because we have used the energy relations only and did not discuss details of the acceleration mechanism. Although, the equation (19) is remarkable. It relates the two lengths, $R$ and $L$. One can introduce the fundamental similarity length $L_{sim} \propto \sqrt{RL} \propto S^{-1} k_0^{-1}$. This length is a consequence of our choice $\varkappa = 0$, which physically means that the laser energy dissipation mechanism is independent on $S$.

Further, substituting Eq. (19) into the first Eq. (18) one obtains $\alpha_\gamma = -1$, i.e., $\gamma = a_0 S^{-1} \gamma_0(\Pi)$. This is the direct consequence of $\alpha_\varkappa = 0$ that corresponds to the most efficient action of the accelerating field $E_{wake}$. The last Eq. in (18) yields $\alpha_N = 3\alpha_R + 1$, i.e., $N \propto n_e R^3$. This is the consequence of our choice $\alpha_\eta = 0$, which optimizes the overall acceleration efficiency.

It follows from the bubble physics [9, 12], that if the aspect ratio $\Pi < 1$, the laser pulse fits into the cavity. In this case, it is reasonable to expect that the efficiencies $\varkappa_0, \theta_0, \eta_0$ weakly depend on $\Pi$. For simplicity, we neglect this weak dependence. Then, the last three equations in (18) claim that the dependence of $N_0$ on $\Pi$ is also weak. At the same time, $\gamma_0(\Pi)$ is simply propotional to $\Pi$. Summarising, we write

$$\gamma \propto a_0 S^{-1} \Pi, \quad N \propto n_e R^3. \quad (21)$$

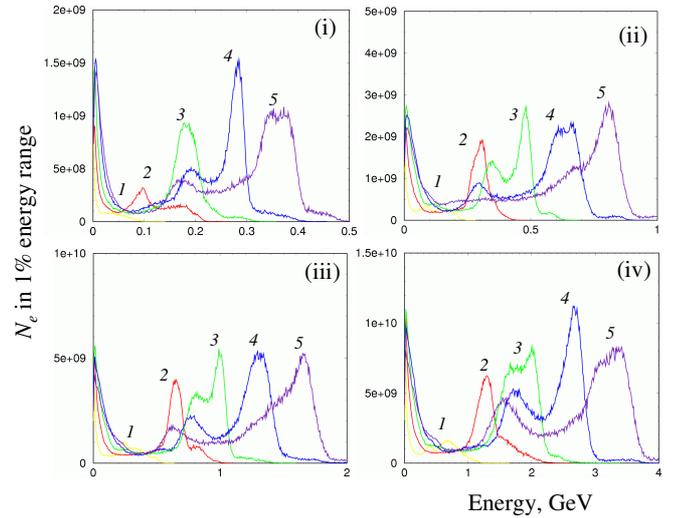

FIG. 1: Electron energy spectra obtained in the simulations (i)-(iv) (see text). The control points $1 - 5$ were taken after the propagation distances $L_1 = 200\lambda$, $L_2 = 400\lambda$, $L_3 = 600\lambda$, $L_4 = 800\lambda$, $L_5 = 1000\lambda$. The spectra evolve similarly. The monoenergetic peak positions scale $\propto a_0$ and the number of electrons in a 1% energy range also scales $\propto a_0$ in agreement with the analytic scalings (11).

To obtain further scalings on the radius $R$ and the length $L$, one needs additional information on the accelerating structure. It follows from the Maxwell equations [12] that the bubble potential $\phi$ scales together with the laser potential $a_0$. Because $\phi \propto nR^2$, we conclude that $k_0 R \propto S^{-1/2}$ and that the dependence $R_0(\Pi)$ is weak. Finally, we obtain for the acceleration length

$$L \propto S^{-3/2} \Pi \propto R^3 \Pi, \qquad (22)$$

and for the similarity length $L_{sim} = \sqrt{RL} \propto R^2$. The presence of these two different lengths leads to the so-called "ladder" similarity as discussed in detail in [12].

Adding dimensional factors to the scalings (21)-(22) and comparing with numerical simulations we come to the formulas (1)-(3). We emphasize once more that these scalings describe the optimal acceleration regime in the limit of small $S$, because we have chosen the largest physically allowed parametric dependencies for the accelerating force and the energy trasformation efficiency.

To check the analytical scalings, we use 3D Particle-in-Cell simulations with the code VLPL (Virtual Laser-Plasma Laboratory) [10]. In the simulations, we use a circularly polarized laser pulse with the envelope $a(t, \mathbf{r}_\perp) = a_0 \cos(\pi t/2\tau) \exp(-\mathbf{r}_\perp^2/R^2)$, which is incident on a plasma with uniform density $n_e$.

First, we check the basic ultra-relativistic similarity with $S = const$. We choose the laser pulse duration $\tau = 8 \cdot 2\pi/\omega_0$. The laser radius is $R = 8\lambda$, where $\lambda = 2\pi c/\omega_0$ is the laser wavelength. The laser pulse aspect ratio $c\tau/R = 1$ in this case.

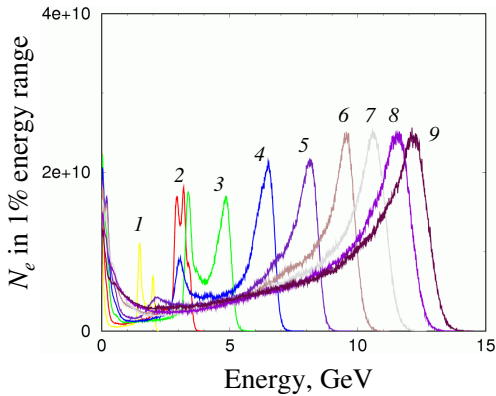

FIG. 2: Electron energy spectra obtained in the simulations (v) (see text). The control points $1-9$ were taken after the propagation distances $L_1 = 800\lambda$, $L_2 = 1600\lambda$, $L_3 = 2400\lambda$, $L_4 = 3200\lambda$, $L_5 = 4000\lambda$, $L_6 = 4800\lambda$, $L_7 = 5600\lambda$, $L_8 = 6400\lambda$, $L_9 = 7200\lambda$. The spectral evolution for the control points $1-5$ is similar to that of the simulation cases (i)-(iv). The spectra $6-9$ correspond to a new evolution that cannot be directly scaled from the previous simulations.

We fix the basic similarity parameter to the value $S^{\text{i}} = 10^{-3}$ and perform a series of four simulations with (i) $a_0^{\text{i}} = 10$, $n_e^{\text{i}} = 0.01 n_c$; (ii) $a_0^{\text{ii}} = 20$, $n_e^{\text{ii}} = 0.02 n_c$; (iii) $a_0^{\text{iii}} = 40$, $n_e^{\text{iii}} = 0.04 n_c$; (iv) $a_0^{\text{iv}} = 80$, $n_e^{\text{iv}} = 0.08 n_c$. Assuming the laser wavelength $\lambda = 800$ nm, one can calculate the laser pulse energies in these four cases: $W^{\text{i}} = 6$ J; $W^{\text{ii}} = 24$ J; $W^{\text{iii}} = 96$ J; $W^{\text{iv}} = 384$ J. These simulation parameters correspond to the bubble regime of electron acceleration [9], because the laser pulse duration $\tau$ is shorter than the relativistic plasma period $\sqrt{a_0}\omega_p^{-1}$. We let the laser pulses propagate the distance $L_b^{\text{i}} = 1000\ \lambda$ through plasma in the all four cases. At this distance, the laser pulses are depleted, the acceleration ceases and the wave breaks.

Fig. 1(i)-(iv) shows evolution of electron energy spectra for these four cases. One sees that the energy spectra evolve quite similarly. Several common features can be identified. First, a monoenergetic peak appears after the acceleration distance $L \approx 200\ \lambda$. Later, after the propagation distance $L \approx 600\ \lambda$, the single monoenergetic peak splits into two peaks. One peak continues the acceleration towards higher energies, while another peak decelerates and finally disappears. Comparing the axises scales in Fig. 1, we conclude that the scalings (11) hold with a good accuracy.

Now we are going to check the general scalings (21)-(22) for the variable $S$-parameter. We choose the laser amplitude $a_0^{\text{v}} = 80$ and the plasma density $n_e^{\text{v}} = 0.02 n_c$. This corresponds to $S^{\text{v}} = 2.5 \cdot 10^{-4}$ and the laser energy $W^{\text{v}} \approx 1.5$ kJ;. In this case, the initial laser radius and duration must be increased by the factor $\sqrt{S^{\text{i}}/S^{\text{v}}} = 2$. Thus, we use the laser pulse with $R^{\text{v}} = 16\lambda$ and $\tau^{\text{v}} = 16 \cdot 2\pi/\omega_0$. This case gives the pure density scaling when compared with the case (iv), or the pure laser amplitude scaling when compared with the case (ii). We let the laser run $L_{acc}^{\text{v}} = 8000\lambda$ through the plasma. At this distance, the energy of the laser pulse is completely depleted and the wave breaks. The change of the depletion length $L_{acc}^{\text{v}}/L_{acc}^{\text{i}} = \left(S^{\text{i}}/S^{\text{v}}\right)^{3/2}$ coincides with the scaling (22).

The electron spectrum evolution obtained in this simulation is shown in Fig. 2. The energy of the monoenergetic peak continuously grows up to some 12 GeV at the end. Between the control points, where the spectra in Fig. 2 have been taken, the laser pulse propagated the distance $L = 800\lambda$. This distance is $S^{\text{i}}/S^{\text{v}} = 4$ times larger than that in the cases (i)-(iv). One sees that the first five electron spectra in Fig. 2 are similar to those in Fig. 1. However, the last four spectra in Fig. 2 are new. This corresponds to the ladder similarity.

Finally, the present work states that the bubble acceleration regime is stable and scalable. Moreover, it corresponds to the optimal scalable regime of wake field acceleration. The ultra-relativistic similarity theory is developed and "engineering" scalings for the electron acceleration are derived.

This work was supported in parts by the Transregio project TR-18 of DFG (Germany) and by RFFI 04-02-16972, NSH-2045.2003.2 (Russia).